\documentclass[prl,twocolumn,a4paper,superscriptaddress,floatfix]{revtex4}
\usepackage{amsmath,graphicx,color}
\usepackage[utf8]{inputenc}

\let\Im\relax\DeclareMathOperator{\Im}{Im}
\let\Re\relax\DeclareMathOperator{\Re}{Re}
\renewcommand{\vec}[1]{\mathbf{#1}}
\newcommand{\eb}{\varepsilon_B}
\newcommand{\ef}{\varepsilon_F}
\newcommand{\ad}{a_\text{2D}}

\begin{document}
\author{Marianne Bauer}
\email{msb50@cam.ac.uk}
\affiliation{Cavendish Laboratory, Cambridge, CB3 0HE, United Kingdom}
\author{Meera M. Parish}
\affiliation{London Centre for Nanotechnology, Gordon Street, London,
  WC1H 0AH, United Kingdom}
\author{Tilman Enss}
\affiliation{Institut f\"ur Theoretische Physik, Universit\"at
  Heidelberg, 69120 Heidelberg, Germany}

\title{Universal equation of state and pseudogap in the two-dimensional
  Fermi gas}

\begin{abstract}
  We determine the thermodynamic properties and the spectral function
  for a homogeneous two-dimensional Fermi gas in the normal state
  using the Luttinger-Ward, or self-consistent T-matrix, approach.
  The density equation of state deviates strongly from that of the
  ideal Fermi gas even for moderate interactions, and our calculations
  suggest that temperature has a pronounced effect on the pressure in
  the crossover from weak to strong coupling, consistent with recent
  experiments.  We also compute the superfluid transition temperature
  for a finite system in the crossover region.  There is a pronounced
  pseudogap regime above the transition temperature: the spectral
  function shows a Bogoliubov-like dispersion with back-bending, and
  the density of states is significantly suppressed near the chemical
  potential.  The contact density at low temperatures increases with
  interaction and compares well with both experiment and
  zero-temperature Monte Carlo results.
\end{abstract}

\maketitle
The formation of fermion pairs and superfluidity of such pairs are
distinct but related phenomena: in weak-coupling BCS theory, both are
predicted to occur at the same temperature $T_c$.  However, a basic
question of many-body physics is how they are related at stronger
coupling and in low dimensions, where quantum fluctuations play a
large role.  While preformed pairs in the normal phase trivially exist
in the strong-coupling Bose limit where one has tightly bound dimers,
it has been argued that pairing above $T_c$ can also occur in the BCS
regime.  In this case, one expects a significant suppression of
spectral weight at the Fermi surface even above $T_c$.  This so-called
pseudogap regime extends up to a crossover temperature $T^*>T_c$, and
its spectral and thermodynamic properties deviate strongly from the
predictions of Fermi-liquid theory \cite{randeriareview}.  Recently,
pairing and superfluidity have been studied in ultracold atomic gases,
which afford accurate control of both the interaction strength and
dimensionality, and allow access to the crossover between the BCS and
Bose regimes \cite{BlochDalibardZwerger}.  In these systems, a
pseudogap can be detected through the suppression of the spin
susceptibility or directly via the spectral function, which is
experimentally accessible by ARPES or momentum-resolved rf
spectroscopy \cite{stewartjin, feld11}.  The possibility of a
pseudogap regime has already been investigated both experimentally and
theoretically in three dimensions (3D)~\cite{chenlevin05, stewartjin}.

\begin{figure} 
  \begin{center}
    \includegraphics[width=\linewidth]{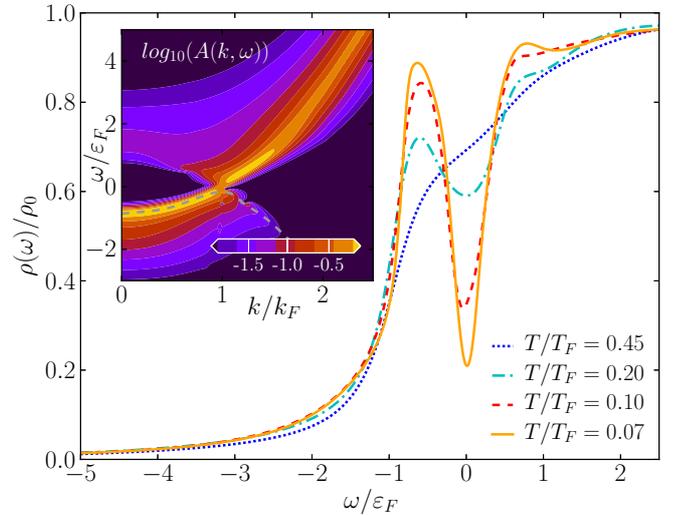}
  \end{center}
  \caption{(Color online) Density of states $\rho(\omega)$, normalised
    by $\rho_0 = m/2\pi$ for the free Fermi gas, at interaction
    $\ln(k_F \ad) = 0.8$ for different temperatures: $T=0.45\,T_F$
    (top curve at $\omega=0$) to $T=0.07\,T_F$ (bottom).
    \textbf{Inset:} Spectral function $A(k,\omega)$ for $T =
    0.07\,T_F$.  The grey dashed line marks the maximum in the
    spectral weight of the bottom band. \label{fig:dos}}
\end{figure}

\begin{figure} 
  \begin{center}
    \includegraphics[width=\linewidth]{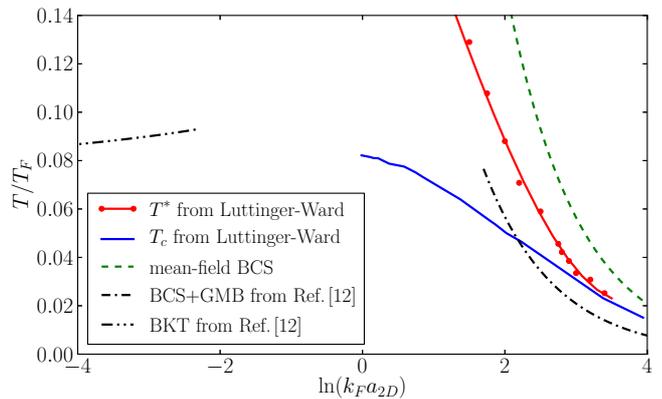}
    \caption{(Color online) Critical temperature $T_c/T_F$ vs
      interaction strength $\ln(k_F\ad)$.  The Luttinger-Ward result
      for a finite system (blue solid line) in the crossover region
      $\ln(k_F\ad)\gtrsim0$ is compared with analytical limits
      \cite{petrov03}.  The red dots marks the crossover temperature
      $T^*$ to the pseudogap regime for $\ln(k_F\ad) \gtrsim
      1$. \label{fig:tc}}
  \end{center}
\end{figure}

In two-dimensional (2D) Fermi gases, the pseudogap regime is expected
to be much more pronounced than in 3D, and a pairing gap has
recently been observed experimentally \cite{feld11}.  Here, we
compute the spectral function for the homogeneous 2D Fermi gas in the
normal phase of the BCS-Bose crossover.  We indeed find a strong
suppression of the density of states at the Fermi surface above $T_c$,
as shown in Fig.~\ref{fig:dos}.  This allows us to map the extent of
the pseudogap regime in the temperature-vs-coupling phase diagram
(Fig.~\ref{fig:tc}), and we find that it extends further than in 3D
\cite{watanabe2013}.

As the binding between fermions increases, the Cooper pairs evolve
into a Bose gas of tightly bound molecules.  Long-range fluctuations
in 2D are so strong that they inhibit superfluid long-range order at
nonzero temperature.  Thus, the 2D Bose gas exhibits a
Berezinskii-Kosterlitz-Thouless (BKT) transition at $T_c>0$ into a
quasi-ordered phase with algebraically decaying correlations
\cite{fisher1988, prokofevsv02, holzmannbaym}.  It is a challenging
many-body problem to precisely characterise the crossover between the
bosonic BKT and fermionic BCS limits, where the composite nature of
the molecules becomes apparent.

\begin{figure}
  \begin{center}
    \includegraphics[width=\linewidth]{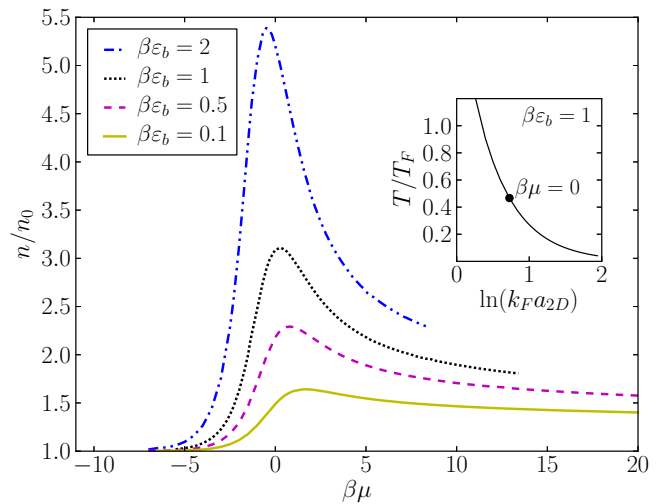}
    \caption{(Color online) Density $n$ of the 2D Fermi gas vs
      chemical potential $\beta\mu$, for different interaction
      strengths $\beta \eb$ (see legend).  Since the density is
      normalised by $n_0(\beta\mu)$ for the non-interacting gas, the
      non-monotonic behavior of $n/n_0$ reflects the impact of
      interactions, while the compressibility $\kappa = (\partial
      n/\partial \mu)/n^2$ is always positive.  The inset shows a
      typical trajectory in $T/T_F$ vs $\ln(k_F\ad)$ corresponding to
      the dotted line of fixed $\beta\eb=1$.  Along this line, $\beta
      \mu$ increases with decreasing $T/T_F$.  \label{fig:density}}
  \end{center}
\end{figure}

\begin{figure} 
  \begin{center}
    \includegraphics[width=\linewidth]{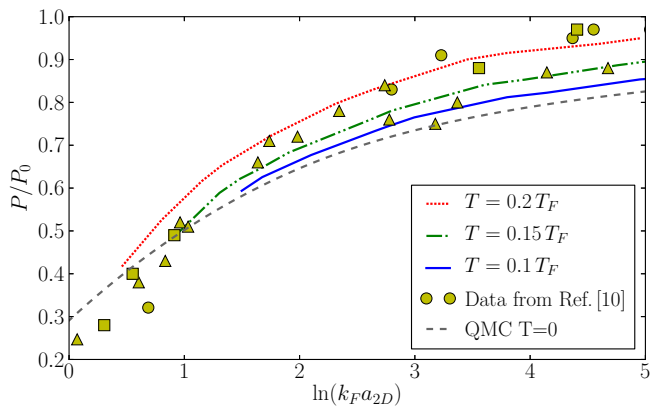}
    \caption{(Color online) Pressure $P$ vs interaction strength,
      normalised by the pressure $P_0 = n\ef/2$ of an ideal Fermi gas
      of the same density at $T=0$.  Luttinger-Ward data at
      temperature $T/T_F=0.2$ (top, dotted) to $T/T_F=0.1$ (solid) in
      comparison with experimental data \cite{makhalov2014} (symbols)
      and $T=0$ Monte Carlo results \cite{bertaina11}
      (dashed). \label{fig:P}}
  \end{center}
\end{figure}

\begin{figure} 
  \begin{center}
    \includegraphics[width=\linewidth]{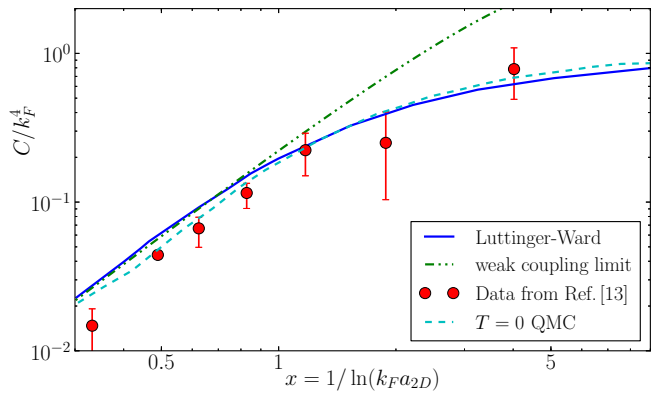}
    \caption{(Color online) Contact density $C$ vs interaction
      strength $1/\ln(k_F\ad)$ at temperature $T/T_F = 0.27$.  We
      compare our result (blue solid line) with the experimental data
      at $T/T_F = 0.27$ \cite{froehlich} (red symbols), the
      weak-coupling result (green dash-dotted line), and Monte Carlo
      at $T=0$ \cite{bertaina11} (cyan dashed
      line).  \label{fig:contact}}
  \end{center}
\end{figure}

In this work, we present the first computation of the
finite-temperature density and pressure equation of state in the
crossover regime and find a strong renormalisation already for
moderate interactions---see Fig.~\ref{fig:density}.  The pressure at
low temperatures has very recently been measured in experiment
\cite{makhalov2014}.  We find that the pressure computed at
$T\simeq 0.1\,T_F$ is closer to the experimental data than zero-temperature
Monte Carlo calculations \cite{bertaina11}, offering a resolution of
previous discrepancies (Fig.~\ref{fig:P}).  Furthermore, we determine
$T_c$ for finite systems (Fig.~\ref{fig:tc}), which is relevant for
experiments on quasi-2D atomic gases, in the crossover regime between
the known limiting cases \cite{petrov03}.  Finally, the contact
density agrees well with experiment \cite{froehlich} and shows
surprisingly little variation with temperature
(Fig.~\ref{fig:contact}).


The dilute, two-component ($\uparrow$, $\downarrow$) Fermi gas with
short-range interactions is described by the Hamiltonian
\begin{align*}
  H = \sum_{\vec{k}\sigma} (\varepsilon_{\vec{k}} - \mu)
  c^\dag_{\vec{k}\sigma}c_{\vec{k}\sigma}
  +\frac{g_0}{V}
  \sum_{\vec{k},\vec{k'},\vec{q}}
  c^\dag_{\vec{k}\uparrow}c^\dag_{\vec{k'}\downarrow}
  c_{\vec{k'}+\vec{q}\downarrow}c_{\vec{k}-\vec{q}\uparrow}
\end{align*}
where $c^\dag_{\vec{k}\sigma}$ creates a fermion with spin $\sigma$,
momentum $\vec{k}$, and kinetic energy $\varepsilon_\vec{k} = \hbar^2
\vec{k}^2 / 2 m$.  The chemical potential $\mu$ is taken to be the
same for both species in a spin-balanced gas.  The energy scale is set
by the Fermi energy $\ef=k_BT_F = \hbar^2k_F^2/2m$ for a total density
$n=k_F^2/2\pi$.  The bare attractive contact interaction $g_0$ has to
be regularised and is expressed in terms of the physical binding
energy $\eb$ of the two-body bound state which is always present in an
attractive 2D Fermi gas.  We define the 2D scattering length as
$\ad=\hbar/\sqrt{m\eb}$ and parametrise the interaction strength by
$\ln(k_F\ad) =\ln(2\ef/\eb)/2$.  In the following we set $k_B = 1$,
$\hbar = 1$, and write $\beta=1/k_BT$.

We investigate the behavior of the strongly interacting Fermi gas in
the normal state using the Luttinger-Ward, or \emph{self-consistent}
T-matrix, approach \cite{haussmann93, haussmann07}, which goes beyond
earlier works \cite{nozieres1985, watanabe2013} by including
approximately the interaction between dimers as well as dressed
Green's functions.  Thermodynamic precision measurements for the
unitary Fermi gas in 3D \cite{ku12} have confirmed the accuracy of
this method, both for the value of $T_c/T_F=0.16(1)$ and the Bertsch
parameter $\xi=0.36(1)$ \cite{haussmann07, ku12}.  Recently, the
Luttinger-Ward approach has been extended to study transport
properties \cite{ensshaussmannzwerger}.  The success of this method in
three dimensions encourages its application to the homogeneous 2D
Fermi gas, which is particularly challenging due to the logarithmic
energy dependence of the scattering amplitude.

Within the Luttinger-Ward approach, pairs of dressed fermions with
Green's function $G(\vec{k}, \omega) = [-\omega + \varepsilon_\vec{k}
- \mu - \Sigma(\vec{k}, \omega)]^{-1}$ can form virtual molecules
whose dynamics are described by the T matrix $\Gamma(\vec{K},
\Omega)$.  The fermions can scatter from these molecules, which
determines their lifetime and self-energy $\Sigma(\vec{k}, \omega)$
(see Supplemental Material \cite{supp}).  From the self-consistent
solution $G(\vec{k}, \omega)$ one obtains the spectral function
$A(\vec{k}, \omega)=\Im G(\vec{k}, \omega+i0)/\pi$.


\textit{Density of states.}---%
The density of states $\rho(\omega)$ describes at which energies
fermionic quasiparticles can be excited, and is computed as the
momentum average of the spectral function, $\rho(\omega)=\int
d\vec{k}\, A(\vec{k},\omega)/(2\pi)^2$.  Figure~\ref{fig:dos} shows
the density of states for an interaction strength of $\ln (k_F \ad) =
0.8$, which is weak enough that there should be a Fermi surface at low
temperatures \cite{wave}.  For decreasing temperature, we see that the
density of states is strongly suppressed at the chemical potential,
while it increases on either side of the Fermi surface.  This marks
the pseudogap regime which is part of the normal phase, but with
anomalous properties due to the lack of low-energy fermionic
excitations.  There is no uniquely defined temperature associated with
this crossover, so for concreteness we define the pseudogap
temperature $T^*$ as the temperature where the density of states at
the chemical potential drops by $25\%$ of the value at the left
fringe.

The full spectral function $A(\vec{k},\omega)$, shown in the inset of
Fig.~\ref{fig:dos} for a temperature of $T/T_F=0.07$ slightly above
$T_c$, shows a BCS-like dispersion with a clear reduction of spectral
weight near the Fermi energy.  While the upper branch has a minimum at
a finite wavevector $k\simeq k_F$, the lower branch exhibits
``back-bending'' towards lower energy for large momenta (cf.\
Ref.~\cite{watanabe2013}).  We note that back-bending alone is not
sufficient to define the pseudogap regime and can arise also for other
reasons in the occupied spectral function \cite{schneider}.  The
two-peak structure of the $k=0$ spectral function qualitatively agrees
with the momentum-resolved RF spectrum measured at $\ln(k_F\ad) = 0.8$
\cite{feld11}, which is the only measurement that may lie within the
pseudogap regime \cite{wave}.  For stronger attraction, the pseudogap
regime eventually crosses over into preformed fermion pairs, where the
Fermi surface is lost ($\mu < 0$) and the spectral function resembles
the one predicted using the virial expansion \cite{wave, barth}.


\textit{Density equation of state.}---%
The total density of both spin components follows from the density of
states as $n=2\int_{-\infty}^\infty d\varepsilon\, f(\varepsilon)
\rho(\varepsilon)$, where $f(\varepsilon)$ is the Fermi distribution.
In Fig.~\ref{fig:density}, we plot the density equation of state
$n(\beta\mu,\beta\eb)$ as a function of $\beta \mu$ for different
values of the interaction parameter $\beta \eb$.  This manner of
plotting the equation of state allows one to make a direct connection
with experiments in trapped gases, since the density versus chemical
potential at fixed $\beta \eb$ can be easily extracted from the
measured density profile in a trap \cite{ku12}.  To expose the effects
of interactions, we normalise the density $n$ by that of the ideal
Fermi gas, $n_0 = 2 \ln(1+e^{\beta \mu})/ \lambda_{T}^2$, where
$\lambda_{T}=\sqrt{2\pi/mT}$ is the thermal wavelength.  In the
high-temperature limit where $\beta \mu \to -\infty$, all properties
approach those of an ideal Boltzmann gas.  However, with decreasing
temperature, we find that $n/n_0$ eventually exhibits a maximum around
$\beta\mu \simeq 0$, implying that interactions are strongest at
intermediate temperatures.  This is easily understood from the fact
that decreasing $T/T_F$ at fixed $\beta\eb$ results in an increasing
$\ln(k_F\ad)$, as shown in the inset of Fig.~\ref{fig:density}.  Thus,
we likewise expect the system to approach a weakly interacting gas in
the low temperature regime.  This behavior is qualitatively different
from that observed in 3D \cite{ku12}, and is a direct consequence of
the fact that one can have a \emph{density}-driven BCS-Bose crossover
in 2D.  The curves for large $\beta\eb$ are shown up to the critical
value $\mu_c(\beta\eb)$ where the system is expected to enter the BKT
phase.


\textit{Pressure.}---%
The pressure is obtained by integrating the density according to the
Gibbs-Duhem relation, $P(\mu)_{T,\eb} = \int_{-\infty}^{\mu} n(\mu')\,
d\mu'$.  Figure~\ref{fig:P} shows the Luttinger-Ward data for finite
temperatures $T/T_F=0.2$ (top) to $0.1$ (bottom): the pressure
decreases from the free Fermi pressure in the BCS limit to the much
lower pressure of a dilute Bose gas in the BKT limit.  This is a
strong coupling effect beyond the mean-field BCS prediction $P=P_0$ at
$T=0$ \cite{bertaina11, makhalov2014}.  As the temperature is lowered,
our data approach the $T=0$ Monte Carlo results \cite{bertaina11}
(dashed).  A recent measurement at low temperatures $T/T_F\simeq
0.04\dotsc 0.12$ \cite{makhalov2014} (symbols) found a deviation from
the $T=0$ pressure in the BCS limit, attributed to mesoscopic effects.
We, however, find that the $T/T_F\simeq 0.1$ pressure from the
Luttinger-Ward calculation agrees well with experiment in this regime,
thus suggesting that the discrepancy is in large part due to the
effect of temperature.


\textit{Phase diagram of the 2D Fermi gas.}---%
The Berezinskii-Kosterlitz-Thouless (BKT) transition at a finite
temperature $T_c$ marks the onset of a nonzero superfluid density
$\rho_s$ and algebraically decaying correlations \cite{fisher1988,
  prokofevsv02}.  The jump in $\rho_s/n$ at $T_c$ is universal for a
Bose superfluid and becomes exponentially small of order $T_c/T_F$ on
the weak-coupling BCS side \cite{stintzing1997}.  The transition
temperature is characterised by the Thouless criterion, where the
coefficient of the quadratic term in a Ginzburg-Landau action for the
pairing field changes sign.  In practice, the relevant question is
when this transition occurs for a finite system, for instance inside a
trapping potential (see Supplemental Material \cite{supp}).  In our
analysis we therefore compute $T_c$ for $N=500$ particles typical of
current experiments \cite{makhalov2014}, as depicted in
Fig.~\ref{fig:tc}.  We have checked that different values for $N$ lead
to small quantitative but not qualitative changes in the $T_c$ curve.

In the weak-coupling BCS limit $\ln(k_F\ad) \gg 1$ [$\eb \ll \ef$],
the mean-field transition temperature is given by $T_c/T_F =
(2e^{\gamma_E}/\pi) \exp[-\ln(k_F\ad)]$ (dashed line in
Fig.~\ref{fig:tc}), where $\gamma_E\approx 0.5772$ is Euler's constant
\cite{miyake83}.  Petrov et al.\ \cite{petrov03} have included
Gor'kov--Melik-Barkhudarov corrections and obtained a lower value
$T_c/T_F = (2e^{\gamma_E}/\pi e) \exp[-\ln(k_F\ad)]$ (dash-dotted
line).  On the BKT side for strong binding $\eb \gg \ef$, the Thouless
criterion fixes $\mu_c = -\eb/2$ and the number equation determines
$T_c$ \cite{sademelo93}.  A more elaborate analysis using Monte Carlo
data for the weakly interacting Bose gas in 2D \cite{prokofevsv02}
yields a BKT temperature of $T_c/T_F \lesssim 0.12$ for $\ln(k_F\ad) <
0$ \cite{petrov03}, which decreases for even stronger binding (left
dashed curve in Fig.~\ref{fig:tc}).  This limiting behaviour implies
the existence of a maximum $T_c$ in the crossover region (cf.\
Ref.~\cite{drechsler1992}), but does not determine its value or the
precise crossover behaviour.

The Luttinger-Ward result for $T_c$ grows monotonically from the BCS
limit towards strong coupling $\ln(k_F\ad)\approx 0$: it suggests a
maximum $T_c$ at negative $\ln (k_F \ad)$, which is unlikely to exceed
$T_c/T_F \lesssim 0.1$.  This is consistent with experiments which did
not observe signatures of superfluidity down to $T/T_F=0.27$
\cite{feld11}, but is considerably lower than a recent calculation for
a harmonically trapped gas \cite{watanabe2013}.

The red dots in the phase diagram in Fig.~\ref{fig:tc} mark the
crossover temperature to the pseudogap regime $T^*$, where the density
of states $\rho(\omega)$ at the chemical potential drops by $25\%$ of
the value at the left fringe.  In the weak coupling BCS limit, $T^*$
approaches $T_c$ since pairing and condensation occur simultaneously,
and both $T_c$ and $T^*$ tend towards the dashed weak-coupling result.
The large pseudogap regime at strong binding leads to clear signatures
in the spin susceptibility and spectral properties well within reach
of current experiments.


\textit{Contact density.}---%
The contact density $C$ \cite{tan08} characterises the probability of
finding particles of opposite spin close to each other
\cite{braaten2008}.  It determines the \emph{universal} high-energy
properties of a quantum gas with contact interactions, e.g., the
momentum distribution function $n_k \to C/k^4$ at large momenta.
The contact density is related to the variation of the pressure
  with scattering length by the adiabatic theorem \cite{tan08large, wave},
\begin{equation*}
  C = -2\pi m \left.\frac{dP}{d\ln\ad}\right\rvert_{\mu,T,V}.
\end{equation*}
Using the weak-coupling expansion of the ground state energy in
$x=1/\ln(k_F \ad)$ \cite{bloom} one obtains at $T=0$: $C = k_F^4 [x^2
- (3/2-2\ln2) x^3]/4$.  In the normal state the contact density
corresponds to the total density of dimers \cite{haussmann09}.

In Fig.~\ref{fig:contact} we show our result for the contact (solid
line) at $T=0.27\,T_F$ and compare with the experimental data at the
same temperature from Fr\"ohlich et al.~\cite{froehlich}, as well as
with the weak-coupling estimate above.  Remarkably, our calculation in
this low-temperature region is very close to the $T=0$ Monte Carlo
result \cite{bertaina11} (dashed line), showing that the contact has
only a weak temperature dependence.  Note, further, that while one
generally expects the contact to decrease with increasing temperature,
our result for larger $\ln(k_F\ad)$ is higher than the contact at
$T=0$ from Monte Carlo, thus suggesting that $C$ is a non-monotonic
function of $T$, similarly to 3D \cite{yu2009}.


In conclusion, we have presented results for the density and pressure
equation of state which shed light on a recent pressure measurement
\cite{makhalov2014}.  The values for the transition temperature $T_c$
and the pseudogap crossover temperature $T^*$ in the phase diagram
reveal a large pseudogap regime; its effect on the spectral function
and low-energy density of states are accessible and relevant for
current experiments using momentum-resolved rf spectroscopy
\cite{feld11}.  We find that the contact depends only weakly on
temperature, providing a robust interaction gauge.  It will be
worthwhile to extend the Luttinger-Ward technique into the
low-temperature BKT phase, which is characterised by binding of
vortex-antivortex pairs, and study the signatures of the superfluid
phase for a trapped 2D Fermi gas.  The BKT transition itself is
revealed by a jump in the sound velocities \cite{ozawa2014}.

We thank Wilhelm Zwerger for suggesting the problem, valuable
discussions and careful reading of the manuscript.  We
acknowledge discussions with Marcus Barth, Stefan Baur, Gianluca Bertaina, Rudolf
Haussmann, Selim Jo\-chim, Michael K\"ohl, Jesper Levinsen, Vudtiwat
Ngampruetikorn, Richard Schmidt and Andrey Turlapov.  MB thanks the Gates Cambridge
Trust for financial support.  MMP acknowledges support from the EPSRC
under Grant No.\ EP/H00369X/2.

\clearpage

\appendix

\section{Supplemental Material: ``Universal equation of state and
  pseudogap in the two-dimensional Fermi gas'' by M.~Bauer,
  M.~M. Parish, and T.~Enss}

\subsection{Luttinger-Ward approach}

The fully dressed fermionic propagator $G(\vec{k}, i\omega)$ is
calculated from the bare propagator $G_0(\vec{k}, i\omega) =
[-i\omega + \varepsilon_\vec{k} - \mu]^{-1}$ via the Dyson equation
\begin{align}
  \label{eq:dyson}
  G^{-1}(\vec{k}, i\omega)
  = G_0^{-1}(\vec{k}, i\omega) - \Sigma(\vec{k}, i\omega),
\end{align}
where $\omega=(2n+1)\pi T$ are fermionic Matsubara frequencies, and
$\Sigma(\vec{k}, i\omega)$ is the self-energy which captures the
interaction effects.

Dilute Fermi gases are well described in the ladder, or T-matrix,
approximation.  The bosonic vertex function is then given by
\begin{align}
  \label{eq:T}
  \Gamma (\vec{K}, i\Omega )
  = [g_0^{-1}(\Lambda) + \chi (\vec{K},i \Omega)]^{-1}
\end{align}
in terms of the bare coupling $g_0(\Lambda)$, which depends on the UV
momentum cutoff $\Lambda$ (see below), and the pair propagator
\begin{align}
  \label{eq:chi}
  \chi (\vec{K}, i \Omega )
  = \int \frac{d\vec{k}}{(2\pi)^2} \frac{1}{\beta} \sum_{\omega} 
  G(\vec{k}, i\omega) G(\vec{K} - \vec{k}, i\Omega - i\omega)
\end{align}
for bosonic Matsubara frequencies $\Omega = 2 n \pi T$.  In the ladder
approximation the bosonic vertex function is equivalent to the T
matrix, which describes the propagation of bound fermion pairs, or
dimers, via dressed fermion-fermion excitations.  Finally, the fermionic
self-energy describes how fermions scatter off (virtual) dimers,
\begin{align}
  \label{eq:sigma}
  \Sigma(\vec{k}, i\omega) 
  = \int \frac{d\vec{K}}{(2\pi)^2} \frac{1}{\beta} \sum_{\Omega}
  G(\vec{K}-\vec{k}, i\Omega-i\omega) \Gamma(\vec{K}, i\Omega).
\end{align}
The pair propagator $\chi$ has a logarithmic ultraviolet divergence
\cite{engelbrechtranderia}; this is regularised by expressing the bare
coupling $g_0(\Lambda)$ in Eq.~\eqref{eq:T} in terms of the physical
binding energy $\eb$ of the two-body bound state which is always
present in an attractive 2D Fermi gas,
\begin{align}
  \label{eq:gtwobody}
  \frac{1}{g_0(\Lambda)}
  = -\int^\Lambda \frac{d\vec{k}}{(2\pi)^2} 
  \frac{1}{2\varepsilon_{\vec{k}} + \eb}.
\end{align}
Note that the existence of a bound state is both necessary and
sufficient for pairing in 2D \cite{randeria1989}.  

Equations~\eqref{eq:dyson}--\eqref{eq:sigma} constitute a set of
coupled integral equations.  The convolution integrals \eqref{eq:chi}
and \eqref{eq:sigma} are computed in Fourier space $(\vec{x},\tau)$ on
a logarithmic grid of $400\times 400$ grid points \cite{haussmann93}
to account for the logarithmically slow decay of the T matrix in 2D.
The integral equations are solved by iteration, and once convergence
is reached one obtains the self-consistent fermion Green's function
$G(\vec{k},i\omega)$ and the T matrix $\Gamma(\vec{K},i\Omega)$,
respectively.  $G(\vec{k},i\omega)$ is analytically continued to real
frequencies $i\omega\to\omega+i0$ using Pad\'e approximants to
determine the spectral function $A(\vec{k},\omega)=\Im
G(\vec{k},\omega+i0)/\pi$.

In the Luttinger-Ward approach, the density is most conveniently
obtained from the Green's function in Fou\-rier space as $n=-2
G(\vec{x}=0,\tau=-0)$ without the need for analytical continuation.
Similarly, in the normal state, the contact density corresponds to the
total density of dimers \cite{haussmann09} and can be expressed in
terms of the self-consistent T matrix as $C = -m^2 \Gamma(\vec{x}=0,
\tau=-0)$.

\subsection{Definition of the 2D scattering length}

The interaction strength in a purely 2D system is characterized by the
physical binding energy $\eb$, see Eq.\ \eqref{eq:gtwobody}, but
different definitions of the scattering length $\ad$ are used in the
literature.  We follow the convention that $\eb=\hbar^2/m\ad^2$, and
hence $\ad=\hbar/\sqrt{m\eb}$ \cite{feld11}.  Alternatively, one may
use $\eb=4\hbar^2/m\ad^2e^{2\gamma_E}$, and consequently
$\ad^\text{alt}=(2/e^{\gamma_E})\hbar/\sqrt{m\eb} =
(2/e^{\gamma_E})\ad$, see e.g.\ \cite{bertaina11, makhalov2014}.  In a
quasi-2D system realized by a harmonic confinement in the third
direction, as is common for ultracold atomic gases, $\eb$ is related
to the confinement length $\ell_z$ and the 3D scattering length $a$,
which can be tuned by a magnetic Feshbach resonance \cite{petrov03,
  BlochDalibardZwerger}.  Ref.~\cite{makhalov2014} argues that the
correct quasi-2D scattering length in the Bose limit is obtained by
matching the scattering amplitudes of the pure and quasi-2D systems.
The interaction parameter $a_2^\text{alt}\sqrt{n_2}$ used in that work
is related to our definition by $k_F\ad = \sqrt\pi e^{\gamma_E}
a_2^\text{alt}\sqrt{n_2}$ for a single-spin density $n_2=k_F^2/4\pi$.

\subsection{Thouless criterion for finite systems}

The transition temperature $T_c$ is characterised by the Thouless
criterion, $\Re \Gamma^{-1}(\vec{K}=0,i\Omega=0)=0$
\cite{haussmann93}.  The T matrix is proportional to the Green's
function $G_\text{bos}$ of a dilute Bose gas as
$\Gamma^{-1}(\vec{K},i\Omega)=(m/4\pi\eb)
G_\text{bos}^{-1}(\vec{K},i\Omega)$ in the BKT limit
\cite{haussmann93}, and the Bose Green's function for an $N$-particle
system with coherence length $\zeta$ approaches
$G_\text{bos}^{-1}(\vec{K}=0,i\Omega=0)=-1/2m\zeta^2 \simeq -T/N$
\cite{holzmannbaym}.  In our analysis we therefore consider the
Thouless criterion $\Re \Gamma^{-1}(\vec{K}=0, i\Omega=0)=
-m/(4\pi\beta\eb N)$ for $N=500$ particles typical of current
experiments \cite{makhalov2014}.

\end{document}